%% file: main.tex
\documentclass[sigconf]{acmart}

\DeclareCaptionLabelFormat{fig_format}{\textbf{Fig #2}}
\graphicspath{{fig/}}

\newcommand{\eg}{{e.g.}}
\newcommand{\etal}{\emph{et al.}}
\newcommand{\eq}{Eq.}

\newcommand{\xxx}[0]{\textsc{SwAN}\xspace}
\usepackage{xspace}
\usepackage{xcolor}
\usepackage{multirow}
\usepackage{enumitem}
\usepackage{makecell}

\AtBeginDocument{%
  }

\copyrightyear{2024}
\acmYear{2024}
\setcopyright{acmlicensed}\acmConference[RecSys '24]{18th ACM Conference on Recommender Systems}{October 14--18, 2024}{Bari, Italy}
\acmBooktitle{18th ACM Conference on Recommender Systems (RecSys '24), October 14--18, 2024, Bari, Italy}
\acmDOI{10.1145/3640457.3688115}
\acmISBN{979-8-4007-0505-2/24/10}

\acmSubmissionID{163}

\begin{document}

\title{Scene-wise Adaptive Network for Dynamic Cold-start Scenes Optimization in CTR Prediction}

\author{Wenhao Li}
\orcid{0000-0002-5428-258X}
\affiliation{%
  \institution{Huazhong University\\of Science and Technology}
  \city{Beijing}
  \country{China}
}
\email{rzliwenhao@hust.edu.cn}

\author{Jie Zhou}
\affiliation{%
  \institution{School of Software,\\Beihang University}
  \city{Beijing}
  \country{China}}
\email{zhoujiee@buaa.edu.cn}

\author{Chuan Luo}\authornote{Corresponding authors.}
\affiliation{%
  \institution{School of Software,\\Beihang University}
  \city{Beijing}
  \country{China}
}
\email{chuanluo@buaa.edu.cn}

\author{Chao Tang}
\affiliation{%
 \institution{Meituan}
 \city{Beijing}
 \country{China}
}
\email{tangchao12@meituan.com}

\author{Kun Zhang}
\affiliation{%
 \institution{Meituan}
 \city{Beijing}
 \country{China}
}
\email{zhangkun32@meituan.com}

\author{Shixiong Zhao}\authornotemark[1]
\affiliation{%
  \institution{The University of Hong Kong}
  \city{Hong Kong}
  \country{China}}
\email{sxzhao@cs.hku.hk}

\renewcommand{\shortauthors}{Wenhao Li et al.}

\begin{abstract}
  \input{sections/abst}
\end{abstract}

  


% \received{20 February 2024}
% \received[revised]{12 March 2024}
% \received[accepted]{July 22 2024}

\maketitle

\section{Introduction} \label{sec:intro}

\input{sections/intro}

\section{Related work} \label{sec:related}

\input{sections/related_work}

\section{Approach} \label{sec:method}

\input{sections/approach}

\section{Experiments} \label{sec:eval}

\input{sections/exp}

\section{Conclusion} \label{sec:conclu}

\input{sections/conclusion}

% \section{Acknowledgments}

% This work was supported in part by the National Key Research and Development Program of China under Grant 2023YFB3307503, in part by the National Natural Science Foundation of China under Grant 62202025, in part by the Young Elite Scientist Sponsorship Program by CAST under Grant YESS20230566, in part by the CCF-Huawei Populus Grove Fund, in part by the Frontier Cross Fund Project of Beihang University, and in part by the Fundamental Research Fund Project of Beihang University.

\begin{acks}
  This work was supported in part by the National Key Research and Development Program of China under Grant 2023YFB3307503, in part by the National Natural Science Foundation of China under Grant 62202025, in part by the Young Elite Scientist Sponsorship Program by CAST under Grant YESS20230566, in part by the CCF-Huawei Populus Grove Fund, in part by the Frontier Cross Fund Project of Beihang University, and in part by the Fundamental Research Fund Project of Beihang University.
\end{acks}

% This section has a special environment:
% \begin{verbatim}
%   \begin{acks}
%   ...
%   \end{acks}
% \end{verbatim}
% so that the information contained therein can be more easily collected
% during the article metadata extraction phase, and to ensure
% consistency in the spelling of the section heading.

% \begin{acks}
% \end{acks}

\newpage
\balance
\bibliographystyle{ACM-Reference-Format}
\bibliography{ref}

% \appendix









\end{document}

%% file: sections/abst.tex
In the realm of modern mobile E-commerce, providing users with nearby commercial service recommendations through location-based online services has become increasingly vital. While machine learning approaches have shown promise in multi-scene recommendation, existing methodologies often struggle to address cold-start problems in unprecedented scenes: the increasing diversity of commercial choices, along with the short online lifespan of scenes, give rise to the complexity of effective recommendations in online and dynamic scenes. In this work, we propose \textbf{S}cene-\textbf{w}ise \textbf{A}daptive \textbf{N}etwork (\xxx
\footnote{\url{https://github.com/ChrisLiiiii/SwAN}}
), a novel approach that emphasizes high-performance cold-start online recommendations for new scenes. Our approach introduces several crucial capabilities, including scene similarity learning, user-specific scene transition cognition, scene-specific information construction for the new scene, and enhancing the diverged logical information between scenes. We demonstrate \xxx's potential to optimize dynamic multi-scene recommendation problems by effectively online handling cold-start recommendations for any newly arrived scenes. More encouragingly, \xxx has been successfully deployed in Meituan's online catering recommendation service, which serves millions of customers per day, and \xxx has achieved a 5.64\% CTR index improvement relative to the baselines and a 5.19\% increase in daily order volume proportion.

%% file: sections/intro.tex
\begin{figure*}
    \centering
    \includegraphics[scale=0.78]{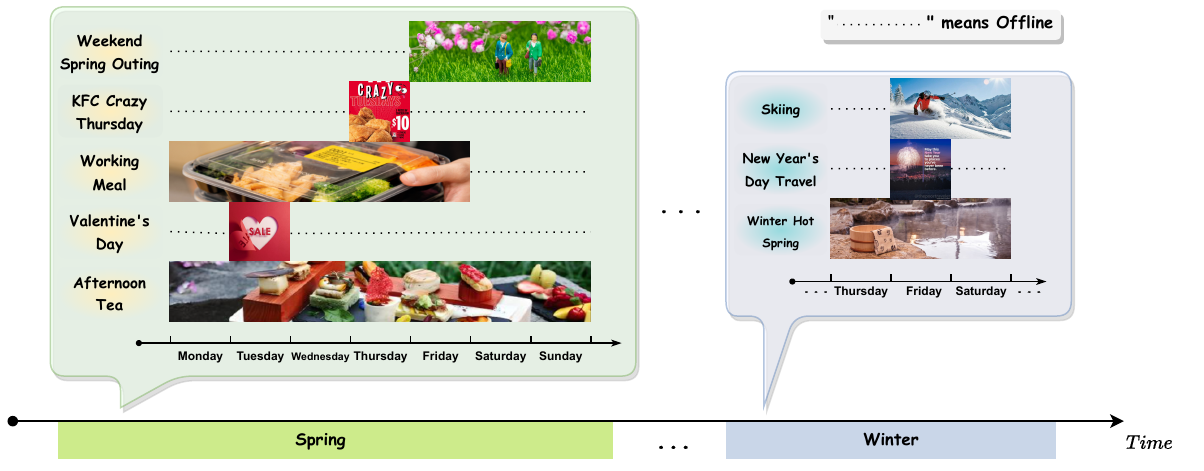}
    \caption{The figure displays multiple business scenes' online / offline states over time. The $x$-axis represents time, with green and blue segments indicating the spring and winter. Boxes above the $x$-axis show the online and offline activities during certain periods within each season. Activity diagrams represent online status, while dashed lines represent offline (e.g., the Valentine's Day activity is only online on Tuesdays in the left graph). The findings suggest that scenes go online immediately when an activity starts (cold-start problem) and go offline right after it ends (limited online time and sample accumulation). The actual online business is even more time-sensitive, with over $200$ scenes going online / offline on average each month. This is the dynamic multi-scene problem introduced in this paper, which poses significant challenges to existing multi-scene models.}
    \label{dynamic_fig}
  \end{figure*}

  Delivering users with nearby commercial service suggestions through location-based online systems \cite{schafer2007collaborative,linden2003amazon,zhou2023feature} has grown increasingly crucial within the era of modern mobile E-commerce.
  Learning to Rank (LTR) involves applying machine learning algorithms~\cite{cheng2016wide,li2023deep} in optimizing the rank strategy, and is the fundamental technique to facilitate better recommendation services. Contemporary recommendation systems not only focus on users' habits derived from historical information but also endeavor to infer the preferences of the same user across diverse scenes, facilitating more accurate and high-quality multi-scene recommendation (MSR)~\cite{zeng2021knowledge}. 

  Despite the considerable advancements in MSR research, a majority of these developments are grounded in the assumption that scenes are predefined and classified prior to offline training, with all subsequent recommendations adhering to established categories during online operations.

  Therefore, the existing literature (\eg SAML~\cite{chen2020scenario}, STAR~\cite{sheng2021one}, and HMoE~\cite{li2020improving}) on MSR primarily concentrates on a static model architecture that distinguishes scenes by directing inputs of each scene to a fixed structural branch within the model.
   
  However, empirical evidence reveals that this assumption does not always hold true. As the assortment of items and options expands in today's world, a proliferation of distinct scenes arises. Consequently, users' behaviors tend to diverge more frequently, leading to an increased variety of scenes without previous identical scenes available for reference in historical data~\cite{hu2022can}. 

  According to Fig.~\ref{dynamic_fig}, the online recommendation service will launch different scenes during specific periods in spring or winter, taking into account user preferences and merchant demands. Additionally, it will also design exclusive activities for specific holidays, such as New Year's Day. On the other hand, scenes often have a limited online lifespan before vanishing (\eg Valentine's Day in Fig.~\ref{dynamic_fig}), leaving no opportunity for a recommendation system to collect data, go offline for fine-tuning, and return online~\cite{rama2019deep}. The Hybrid of implicit and explicit Mixture-of-Experts (HMoE)~\cite{li2020improving} demonstrates that the performance of one scene can be enhanced (through training) by the prediction of other scenes. Unfortunately, HMoE still requires learning the historical data of a new scene and sharing information between scenes through re-parameterization.
  
  In this paper, we demonstrate that the performance of a newly-arrived scene can be directly and significantly improved through online prediction using our \textbf{S}cene-\textbf{w}ise \textbf{A}daptive \textbf{N}etwork (\xxx) model. This suggests that cold-starting new scenes is not only feasible but also surpasses the recommendation performance of existing approaches on known scenes.
  
  In general, \xxx employs the typical Embedding\&MLP (Multilayer Perceptron) paradigm for the recommendation~\cite{zhang2019deep} (Sec.~\ref{sec:method}). In the Embedding part, \xxx utilizes the Scene Relation Graph (SRG) to capture graph-structured similarities between scenes based on inherent attributes and user interaction features, thereby learning the inertial patterns among scenes. The \xxx model also incorporates the Similarity Attention Network (SAN) to capture users' habits during scene transitions by applying user attention on scene similarity knowledge. Furthermore, \xxx assigns each known scene a separate feature embedding (Scene Embedding Layers) to understand how scenes individually influence user behavior and interlace them with the SAN, allowing the impact of new scenes on users to be directly derived. In the MLP part, \xxx generally adopts the Adaptive Ensemble-experts Module (AEM), which is a Mixture-of-Experts (MoE) architecture and includes an Adaptive Expert Group (AEG) of Sparse MoE that uniquely leverages Cosine Loss to enhance diversities between scenes, as well as a Shared Expert Group (SEG) of Multi-gate MoE that captures the shared logic of scenes. In particular, a novel component named $Dics$ has been proposed for the AEG to achieve gradient propagation and select appropriate model structures adaptively.
  
  Extensive evaluation on both the public and industrial datasets shows that the \xxx model outperforms existing MSR approaches by seamlessly adapting to new scenes and providing more accurate and high-quality recommendations. \xxx achieves up to 5.64\% online CTR improvement relative to the baselines and up to 5.19\% increase in daily order volume proportion, as evaluated in Sec.~\ref{app_in_prac}.
  
  The main contributions of this paper are as follows:
  \begin{itemize}
  \item We propose \xxx, an innovative high-performance multi-scene cold-start optimization network.
  \item Innovatively, we propose SRG to acquire prior information from similar scenes for cold-start scenes and employ SAN to get the attention weight of these scenes from user's perspective. Finally, AEM dynamically allocates model structures to enhance the extraction capability of shared and specific information across different scenes.
  \item \xxx has been deployed in a real-world online business recommendation system of Meituan and achieved a 5.64\% improvement in CTR compared to the baseline model.
  \end{itemize}

%% file: sections/related_work.tex
\begin{figure}[t]
    \centering
    \includegraphics[scale=0.70]{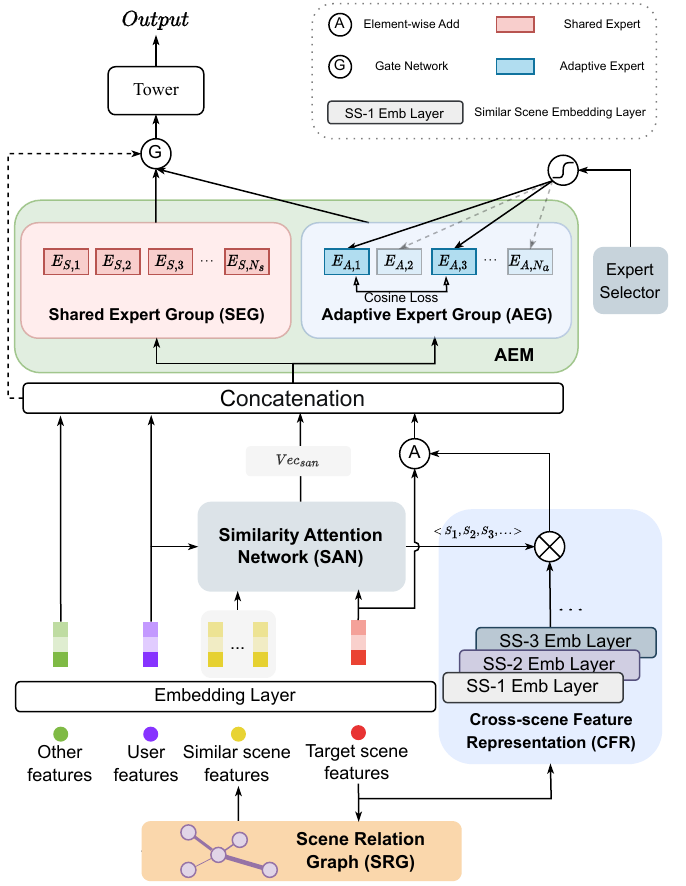}
    \caption{Schematic diagram of the \xxx structure.}
    \label{backbone}
\end{figure}

Multi-scene learning tackles recommendations for users across various scenes~\cite{zhu2021cross}. Traditional models for multi-scene learning have been developed to enhance performance in multiple fixed scenes. Drawing inspiration from the Multi-task Mixture-of-Experts model, Li~\cite{li2020improving} introduced HMoE that implicitly identifies scene disparities and similarities in the feature space and explicitly enhances performance in the label space using a stacked model.

Unfortunately, the model did not distinguish different scenes expert-wise, resulting in insufficient mining of scene-specific information and affecting the model's ability to represent scenes.

To capture the diverse characteristics of various scenes and thus serve them equitably, Sheng \etal~\cite{sheng2021one} brings the STAR model which leverages data from all scenes. PEPNet~\cite{chang2023pepnet} implements an efficient, low deployment cost, plug-and-play multi-scene modeling paradigm by constructing parameters and embedding personalization. Zhou et al.~\cite{zhou2023hinet} proposed HiNet, which employs multi-tasks and multi-scenes explicitly and hierarchically using a hierarchical MoE to model commonality and individuality among multi-scenes.

However, the aforementioned models are designed for multiple fixed scenes, so they need re-training with additional model structure when applied to dynamically increasing cold-start scenes.

The cold-start problem is an open and challenging research problem in the field of recommendation systems~\cite{gope2017survey}. 

Zhu \etal~\cite{zhu2021learning} proposed the Meta Warm Up Framework (MWUF) based on meta-learning and considered that the embeddings for cold-start and warm-up stages are in different spaces. The MWUF designed Meta Scaling Network and Meta Shifting Network to map cold-start embeddings to the warm-up space and eliminate noise. However, the MWUF mainly optimizes item cold-start and is unsuitable for scene cold-start. Besides, Du \etal~\cite{du2019sequential} developed the scene-specific Sequential Meta learner ($s^{2}Meta$) based on meta-learning. The $s^{2}Meta$ model mainly learns the gradient and loss changes of the target model while fitting the distribution of old scene data through the Long Short-Term Memory (LSTM)~\cite{hochreiter1997long} module to guide the model's training direction and early-stop timing in new scenes. Nevertheless, it results in high-cost consumption when applied to multiple cold-start scenes, and it cannot comprehensively consider the distribution rules of multiple scene data.

%% file: sections/approach.tex
  This section presents our design of the proposed \xxx model (Fig.~\ref{backbone}). In essence, \xxx follows the key principle of optimizing multi-scene (by extracting scene-specific and shared information) and cold-start (by incorporating data from similar scenes as supplements) problem and consists of multiple modules: the Scene Relation Graph (Sec.\ref{sec:method:srg}), Similarity Attention Network (Sec.\ref{sec:method:san}), Cross-scene Feature Representation (Sec.\ref{sec:method:asm}), Adaptive Ensemble-experts Module (Sec.\ref{sec:method:mfn}). The Decision Layer (Sec.\ref{sec:method:output}) of \xxx and the loss function (Sec.\ref{sec:method:loss}) are appended.

  \subsection{Scene Relation Graph (SRG)} \label{sec:method:srg}

  A scene comprises inherent attribute features and user interaction features~\cite{chang2023pepnet}. In dynamic multi-scene problems, there is no historical interaction data between users and new scenes, which means that only scene attribute features can be invoked to collect information.
  
  Fortunately, users exhibit similar preferences in comparable scenes. Based on our post-fact online business analysis, users often perceive a positive correlation between the similarity of scenes and the similarity of item features within those scenes~\cite{gu2021self}. This allows a recommendation system to optimize the cold-start process by leveraging prior information from analogous scenes to resemble the target scene closely.
  
  \begin{figure}
      \centering
      \includegraphics[scale=0.58]{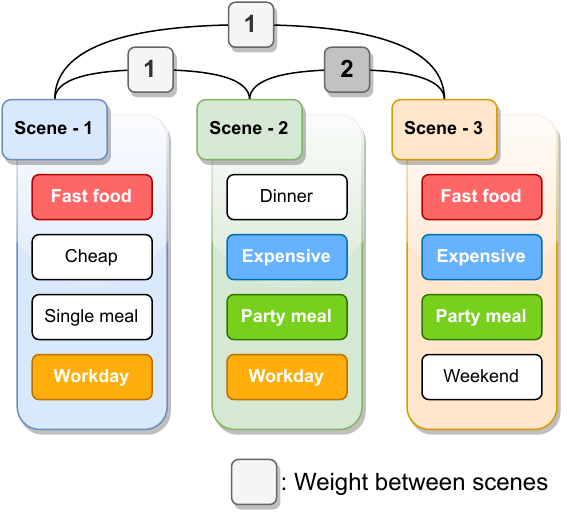}
      \caption{The relation between different scenes in SRG. The numbers on the lines are the number of the same key features.}
      \label{scene_graph}
  \end{figure}
  
  Based on these premises, \xxx invokes a Scene Relation Graph (SRG) module that builds a relational graph between the current scene (to be predicted) and the existing scenes based on the scene features. The construction process of SRG is as follows:
  \begin{enumerate}
  \item Firstly, it lists the basic features (unrelated to online interactions, \eg price and category) of the items to be sorted and uses user key interactions as labels to calculate the Pearson correlation coefficient of various features in the existing scenes, selecting the top-$n$ (Sec.~\ref{hp_srg_cc} for details) key features.
  \item Secondly, it aggregates the key features of items in cold-start scenes to obtain scene-level features such as averages, variances, maximums, and minimums (measuring the distribution patterns of each feature).
  \item Lastly, it categorizes the above features and counts the number of identical features between scenes as the edge weights (Fig.~\ref{scene_graph}, Scene-2 and Scene-3 share 2 identical attributes).
  \end{enumerate}
  
  By doing so, the SRG module obtains a similarity rank between the current and existing scenes by calculating raw feature explicit similarity, making \xxx flexible in choosing a threshold to invoke scenes with certain weighted similarities to the target scene.

  \subsection{Similarity Attention Network (SAN)} \label{sec:method:san}
  
  \begin{figure}[t]
      \centering
      \includegraphics[scale=0.75]{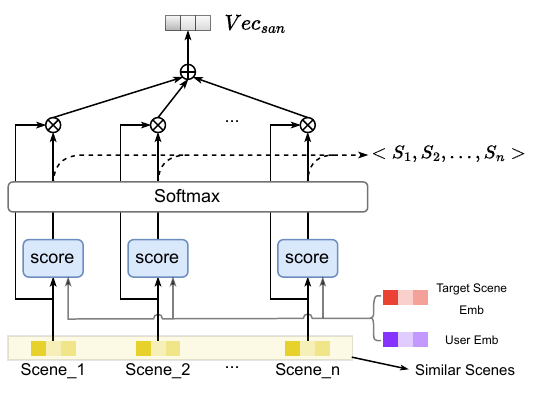}
      \caption{Similarity Attention Network.}
      \label{san_fig}
  \end{figure}
  
  However, determining similar scenes based solely on attributes is inadequate. In real-world applications, various users perceive the same scene pair differently, and the model must incorporate user cognition to comprehend the latent similarity between scenes on a deeper level~\cite{zhu2022personalized}. For instance\footnote[2]{Overall, from our billion-level dataset, we found that 71.69\% of users prefer breakfast within 2 km, while 63.41\% of users choose regular meals within a 2-5 km range. Hence, the distance feature has varying impacts on breakfast and regular meal recommendations. Regarding Sec.~\ref{sec:method:asm}, from our billion-level labeled data, only 49.62\% of users consider zombie movies as horror films, and 35.42\% express opposing views, indicating diverse perceptions among users (zombie movies can trigger aversion in some users). Additionally, the data analysis of various features, such as price, supports our findings.}, some individuals consider horror movies and zombie movies part of the same genre, while others do not. 

  Consequently, our model enhances the SRG module by introducing user information for attention. This is achieved by incorporating a Similarity Attention Network (SAN, shown in Fig.~\ref{san_fig}) to calculate learned latent similarity from the user's perspective.

  The input of the SAN includes the features of the target scene, the similar scenes defined in the SRG, and the user. The specific attention calculation (referred to the DIN~\cite{zhou2018deep}) is as follows:
  \begin{equation}
    \hat{S}_{i} = MLP[E_{u} \oplus E_{t} \oplus (E_{t} - E_{s}^{i}) \oplus (E_{t} \otimes E_{s}^{i})] ,
    \label{san_equ_1}
  \end{equation}
  \begin{equation}
    S_{i} = softmax(\hat{S}_{i}) ,
    \label{san_equ_2}
  \end{equation}
  \begin{equation}
    Vec_{san} = {\sum^{I}} S_{i} \cdot E_{s}^{i} ,
    \label{san_equ_3}
  \end{equation}
  where $E_{u}$, $E_{t}$, and $E_{s}^{i}$ are the embeddings of users, target scene, and the $i$-th similar scene, respectively; $I$ denotes all the scenes; $\hat{S}_{i}$ is the intermediate variable (The output of the blue "score" module in Fig.~\ref{san_fig}.); $S_{i}$ is the latent learned similarity between the $i$-th similar scene and the target scene; operator $\oplus$ means concatenation, and operator $\otimes$ means element-wise product. This structure effectively integrates user cognition to learn the genuine similarity between scenes and outputs a weighted representation of prior information from analogous scenes, which enables a more rational ranking of samples in new scenes. Furthermore, the SRG and SAN structures boosted the performance of \xxx during the cold-starting of new scenes.

  \subsection{Cross-scene Feature Representation (CFR)} \label{sec:method:asm}
  
  There are differences in the bottom-level feature representation for each scene as well~\cite{chen2020scenario}. For example, the distance between users and dining locations has different importance in the breakfast and regular meal scenes\footnotemark[2]. 

  To reflect these differences and provide information supplementation for the cold-start embedding of the target scene, \xxx added a Cross-scene Feature Representation (CFR) structure to the feature processing module (Fig.~\ref{backbone}), which essentially assigns each extent scene a separate embedding to capture a scene's properties solely. Specifically, the input of CFR is the scene-related features $f_{target\_scene}$ and the similarity between scenes output by SAN, and the calculation formula is as follows:
  \begin{equation}
    E_{cfr} = \sum^{I} S_{i} \cdot EMB_{i}(f_{target\_scene}) ,
    \label{asm_equ}
  \end{equation}
  where $EMB_{i}(\cdot)$ is the embedding layer corresponding to the $i$-th similar scene. The input of the subsequent model is:
  \begin{equation}
    E_{in} = E_{o} \oplus E_{u} \oplus Vec_{san} \oplus (E_{t} + E_{cfr}) ,
    \label{input_equ}
  \end{equation}
  where $E_{o}$ means the embedding of other features, and $+$ means element-wise addition.% (\circled{A} in Fig.~\ref{backbone}).
  
  The model transfers prior information from similar scenes regarding feature representation dimensions through CFR, optimizing the cold-start problem and enhancing the expression of differences between scenes. In addition, since CFR essentially involves multiple dictionary lookups and weighted vector summation, it does not introduce excessive computational overhead.

  \subsection{Adaptive Ensemble-experts Module (AEM)} \label{sec:method:mfn}

  Traditional static multi-scene models usually set up separate model branches for each scene (\eg STAR~\cite{sheng2021one}), using structural differences to improve the ability to mine diverged information and optimize negative transfer problem, which are the cores of multi-scene modeling. However, in dynamic multi-scene problems, numerous scenes go online and offline frequently. The traditional model design approach cannot assign model structure for cold-start scenes, while the strategy of retraining the model based on a small number of cold-start scene samples leads to computational redundancy and require frequent offline fine-tuning to update the model architecture. To solve the above problems, we designed Adaptive Ensemble-experts Module (AEM) as the backbone network of the model to enhance the ability to extract differential information and optimize negative transfer in dynamic and multi-scene environments (Fig.~\ref{backbone}).
  
  Firstly, we draw inspiration from the MMoE model~\cite{ma2018modeling} and develop multiple expert networks to enhance the model's ability to mine information. Secondly, we divide the expert networks into groups that improve the model's ability to extract scene-specific and shared information.
  
  The Adaptive Experts Group (AEG) is responsible for extracting scene-specific information. To avoid the high cost of model training caused by frequent scene updates, AEG adopts a dynamic combination of experts to calculate differentiated weights for different scene samples, which means learning how to allocate model structures adaptively. This function is mainly implemented by the Expert Selector (ES). As shown in Fig.~\ref{backbone}, the input of ES is the weighted similar scene representation obtained by SAN, and the output is the gate weight (0 or 1) of each expert in AEG. In this way, ES transfers the prior information of expert selection from similar scenes. AEG enhances the model's ability to extract different scene-specific information and optimizes the cold-start phase of new scenes.
  
  In more detail, the computations in ES consist of two steps: (1) Generate selection probabilities for each expert and the unique threshold (to unify all expert-selected baselines and enhance stability). (2) Output a weight of 1 if the probability exceeds the threshold and 0 otherwise. The specific formulas are as follows:
  \begin{equation}
    P_{k} = sigmoid[ MLP_{p}(E_{u} \oplus Vec_{san}) ] ,
    \label{fn_equ_1}
  \end{equation}
  \begin{equation}
    T = sigmoid[ MLP_{thre}(Vec_{san}) ] ,
    \label{fn_equ_2}
  \end{equation}
  where $P_{k}$ is the selection probability of the $k$-th expert, and $T$ is the probability threshold. In addition, incorporating $E_{u}$ into the calculations of $P_{k}$ allows for more accurate computations from the user's perspective, similar to the SAN.
  
  However, using a step function or a threshold function to compare probabilities and thresholds can lead to gradient interruption, which means that the MLP model for generating probabilities and thresholds cannot be trained. To solve this problem, we have designed a Differentiable conditional selection unit ($Dics$) based on the sigmoid function:
  \begin{equation}
    W_{k} = Dics(P_{k}, T) = \frac{1}{1 - e^{-\frac{1}{\tau} \cdot (P_{k} - T)}},
    \label{dics}
  \end{equation}
  where $\tau$ is a temperature coefficient greater than $0$ and $W_{k}$ is the weight of the $k$-th expert. By performing the aforementioned operations, $Dics$ dynamically selects appropriate model structures for cold-start scenes based on similar scene information, achieving adaptability at the model architecture level.
  
  When the value of $\tau$ is close to $0$, the output weight is more relative to $0$ or $1$. However, excessively small $\tau$ leads to unstable model training. To address this issue, we introduce the variance loss of the gate values for each expert to increase the variance between gate values and move them closer to 0 and 1.
  
  AEG is built on ensemble learning, so improving the differences between sub-learners' outputs can help enhance model performance~\cite{sagi2018ensemble}. We add a cosine similarity loss function between the outputs of each expert in AEG, with the specific formula as follows:
  \begin{equation}
    Loss_{cos} = \sum_{m}^{AEG} \sum_{n \ne m}^{AEG} \left | \frac{E_{a}^{m} \cdot E_{a}^{n}}{\left \| E_{a}^{m} \right \| \cdot \left \| E_{a}^{n} \right \| } \right | ,
    \label{cosine_loss}
  \end{equation}
  where $E_{a}^{m}$ and $E_{a}^{n}$ are output vectors of two different experts in the AEG, and $AEG$ is the whole AEG.
  
  In addition, \xxx constructs the Shared Experts Group (SEG) to enhance the extraction performance of shared information between scenes, following the approach of classic multi-scene models\cite{cheng2016wide}. The experts of this module remain consistent across all scenes.
  
  In summary, AEM adaptively transfers the model-building approach for similar scenes to the current one. The model enhances the generalization ability to new scenes and improves the ability to mine scene-specific and shared information, providing a solution to optimize dynamic multi-scene problems.

  \subsection{Decision Layer} \label{sec:method:output}
  
  By utilizing AEM, \xxx extracts the shared and specific information of scenes, which is contained in the output vectors of each expert in SEG and AEG, respectively. However, the contribution of each vector to the final prediction target varies. To address this issue, inspired by the solution of MMoE, we add a gating network for each expert:
  \begin{equation}
    G_{i} = MLP_{g}(E_{in}) ,
    \label{gate_equ_1}
  \end{equation}
  \begin{equation}
    E_{final\_in} = \sum_{i=0}^{|SEG|} G_{i} \cdot Vec_{s}^{i} + \sum_{i=0,k=0}^{|AEG|} G_{i} \cdot W_{k} \cdot Vec_{a}^{i} ,
    \label{gate_equ_2}
  \end{equation}
  where $E_{final\_in}$ is the input of the MLP structure in the output stage of \xxx, $G_{i}$ is the gate value of each expert, $MLP_{g}(\cdot)$ is the MLP structure to calculate the gate value, and $Vec_{s}$ and $Vec_{a}$ are the expert outputs of SEG and AEG, respectively. The output of the decision layer, namely, the final output of \xxx, can be expressed as:
  \begin{equation}
    Output = sigmoid[MLP(E_{final\_in})] .
    \label{output_equ}
  \end{equation}

  \subsection{Composition of Losses}\label{sec:method:loss}
  
  \xxx uses the Cross-Entropy loss function between output and label to guide training. The formula is as follows:
  \begin{equation}
    Loss_{ce}(y,\hat{y}) = \frac{1}{N} \sum -[y \cdot log(\hat{y}) + (1-y) \cdot log(1-\hat{y})] ,
    \label{loss_ce}
  \end{equation}
  where $y$ and $\hat{y}$ are label and predicted value, respectively.
  
  In addition, as described in Sec.~\ref{sec:method:mfn}, a variance loss is added:
  \begin{equation}
    Loss_{var} = \frac{\sum_{A}(W - \bar{W})^{2} }{N_{a}} ,
    \label{loss_var}
  \end{equation}
  where $N_{a}$ is the number of experts in AEG.
  
  To sum up, the loss function is as follows:
  \begin{equation}
    Loss = \alpha \cdot Loss_{ce}(y, \hat{y}) + \beta \cdot Loss_{cos} + \gamma \cdot Loss_{var} ,
    \label{loss}
  \end{equation}
  where $\alpha$, $\beta$, and $\gamma$ are hyper-parameters set according to the actual dataset (Sec.~\ref{hyper_p_exp}).

%% file: sections/exp.tex
To verify the effectiveness and generalization of the \xxx model, this study conducts experiments based on two datasets: a closed-source dataset from Meituan's online catering recommendation service with millions of daily users and an open-source dataset constructed from the Taobao public dataset~\cite{du2019sequential}.

\subsection{Experimental Settings}

\begin{table}[h]
  \centering
  \caption{The number of scenes contained in the dataset.}
  \small
  \label{dataset_intro}
  \begin{tabular}{c|c|c|cc}
  \hline
  \multirow{2}{*}{Dataset} & \multirow{2}{*}{Source} & \multirow{2}{*}{Train} & \multicolumn{2}{c}{Test}                  \\ \cline{4-5} 
                           &                         &                        & \multicolumn{1}{c|}{Overall} & Cold-start \\ \hline\hline
  1                        & Meituan                 & 751                    & \multicolumn{1}{c|}{309}     & 207        \\
  2                        & Taobao                  & 250                    & \multicolumn{1}{c|}{105}     & 105        \\ \hline
  \end{tabular}
\end{table}

\textbf{Industrial Dataset.} 
Samples from Dataset-1 are obtained from the online catering recommendation platform of
Meituan, specifically from the business of Sales Campaign Session with an average daily customer level of millions. This business designs promotional activity scenes to cater to the consumption preferences of users at different periods, thus involving many scenes and frequent updates. We selected three months and one month of actual user online interaction behaviors as the training and test sets, respectively. The training set contains 70 million samples, while the test set contains 30 million. The ratio of the number of positive samples between the number of negative samples is about 1 to 3. Table~\ref{dataset_intro} shows that this dataset's training and testing sets contain $751$ and $309$ scenes, respectively. Among them, $207$ scenes in the testing set have never appeared in the training set, representing cold-start scenes. 

\begin{table}[ht]
  \caption{Experimental Results (DMSM: dynamic multi-scene model).}
  \small
  \centering
  \begin{tabular}{c|c|cc|c}
  \hline
  \multirow{3}{*}{Model}           & \multirow{3}{*}{Type} & \multicolumn{2}{c|}{Dataset-1}                     & Dataset-2   \\ \cline{3-5} 
                                   &                       & \multicolumn{1}{c|}{\thead{AUC \\ \textit{all}}} & \thead{AUC \\ \textit{cold-start}} & \thead{AUC \\ \textit{cold-start}} \\
  \hline\hline
  \textbf{\xxx} (Ours)             & DMSM                  & \multicolumn{1}{c|}{\textbf{0.7860}}    & \textbf{0.7799}           & \textbf{0.6733}           \\
  DNN~\cite{covington2016deep}     & SSM                   & \multicolumn{1}{c|}{0.7646}             & 0.7568                    & 0.6601                    \\
  DCN~\cite{wang2017deep}          & SSM                   & \multicolumn{1}{c|}{0.7718}             & 0.7642                    & 0.6617                    \\
  xDeepFM~\cite{lian2018xdeepfm}   & SSM                   & \multicolumn{1}{c|}{0.7741}             & 0.7649                    & 0.6631                    \\
  DCN-v2~\cite{wang2021dcn}        & SSM                   & \multicolumn{1}{c|}{0.7758}             & 0.7655                    & 0.6643                    \\
  MMoE~\cite{ma2018modeling}       & SSM                   & \multicolumn{1}{c|}{0.7755}             & 0.7656                    & 0.6638                    \\
  PLE~\cite{tang2020progressive}   & SSM                   & \multicolumn{1}{c|}{0.7787}             & 0.7701                    & 0.6682                    \\
  HMoE~\cite{li2020improving}      & SMSM                  & \multicolumn{1}{c|}{0.7749}             & 0.7644                    & 0.6673                    \\
  STAR~\cite{sheng2021one}         & SMSM                  & \multicolumn{1}{c|}{0.7731}             & 0.7607                    & 0.6669                    \\
  PEPNet~\cite{chang2023pepnet}    & SMSM                  & \multicolumn{1}{c|}{0.7767}             & 0.7659                    & 0.6679                    \\
  HiNet~\cite{zhou2023hinet}       & SMSM                  & \multicolumn{1}{c|}{0.7759}             & 0.7648                    & 0.6677                    \\ \hline
  \end{tabular}
  \label{experiment}
\end{table}

\textbf{Public Dataset.}
Samples from Dataset-2 are obtained from user click logs of cloud-based theme scenes on Taobao~\footnote[3]{\url{https://tianchi.aliyun.com/dataset/9716}}. We followed the official instructions~\cite{du2019sequential} and divided the dataset into training and testing sets, which include $250$ and $105$ different recommendation scenes, respectively.
Actually, after dealing with Dataset-2 through the official instructions, it is guaranteed that none of the testing set scenes appears in the training set. In addition, to adapt the features of the Taobao dataset to the cold-start multi-scene recommendation case studied in our paper, we performed data clustering and further processing of the Taobao dataset.

\begin{figure}
  \centering
  \includegraphics[scale=0.25]{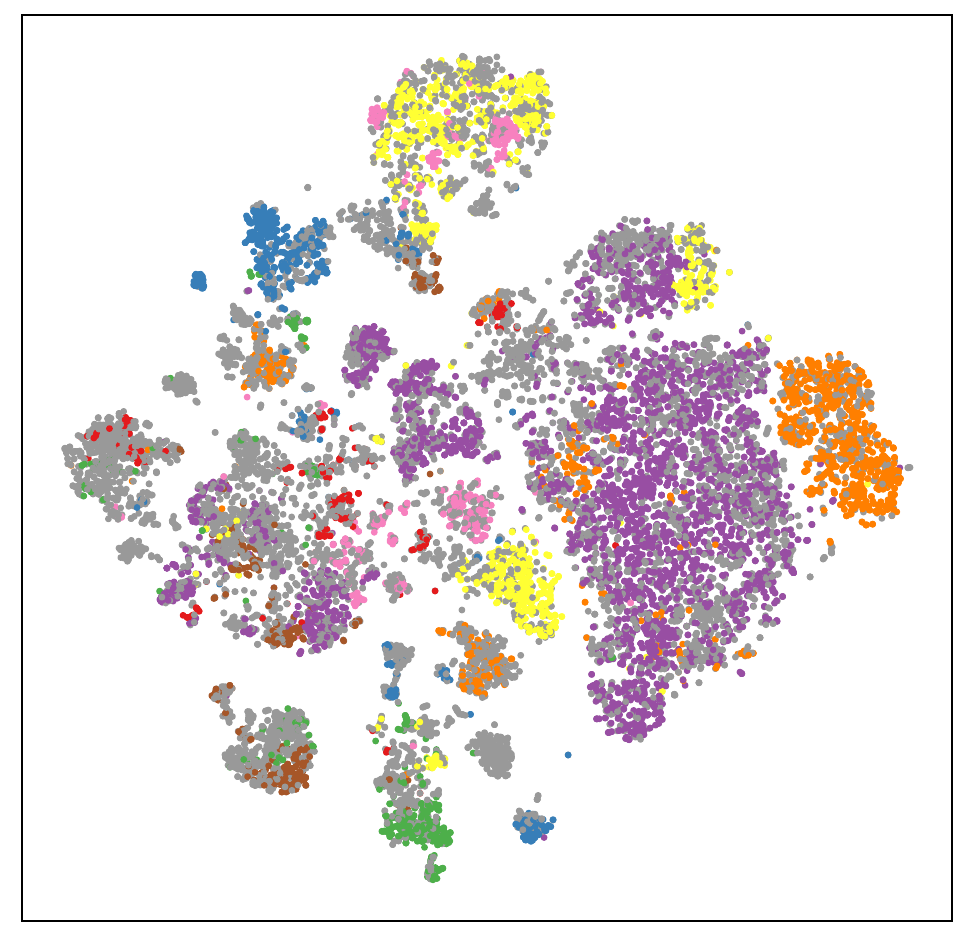}
  \caption{The t-SNE dimensionality reduction visualization of item embeddings, where different colors represent different categories in the original data (gray represents missing category information).}
  \label{tsne_fig}
\end{figure}

The original Taobao dataset only contains item embeddings, scene theme IDs, and some item categories. The original dataset is not well suited for an effective evaluation of \xxx due to two rationales. First, the coverage of item category features is deficient (only 23.18\%), which cannot produce practical scene attributes. Second, when we use t-SNE to reduce the dimensionality of item embeddings (as shown in Fig.~\ref{tsne_fig}), it can be observed that the distribution of items on the two-dimensional plane is not optimal (different colors represent different original categories). Specifically, there are cases where different categories of items are clustered together, and items of the same category are dispersed. This graph also indicates a significant dissimilarity between different original categories. Therefore, it is unreasonable to calculate scene features using the item category information from the original data.

\begin{table}[h]
\centering
\small
\caption{Silhouette coefficients corresponding to different $k$ values. The Silhouette coefficient measures how well each data point fits within its cluster and how well separated it is from other clusters. }
\label{sihouette_coefficient}
\begin{tabular}{c|c|c|c|c|c}
\hline
$k$                    & 2      & 3               & 4      & 6      & 9      \\ \hline\hline
Silhouette Coefficient & 0.1233 & \textbf{0.1437} & 0.1387 & 0.1160 & 0.0993 \\ \hline
\end{tabular}
\end{table}

In summary, the information in the original dataset does not meet the case setting of this paper. To address this problem, we applied the k-means algorithm to cluster the item embeddings provided in the dataset. We used the silhouette coefficient to evaluate the appropriateness of the selected hyper-parameter $k$. The relationship between $k$ and the silhouette coefficient is shown in Table~\ref{sihouette_coefficient}. The table shows that the optimal value for $k$ is $3$.
Using clustering of item embeddings, we obtained the category information for all items. Further, we derived the inherent attributes of each scene, which meets the data requirements of our model in this paper. 

\textbf{Settings.} The cross-entropy loss function and Adam optimizer~\cite{kingma2014adam} are used in the experiments. The number of experts in AEG and SEG is set equally to $10$ as the default value. The value of $\tau$, which is used in the $Dics(\cdot)$ function (Eq.~\ref{dics}), was set to $10^{-3}$ as default. Furthermore, we set $\alpha$ as $1$, $\beta$ and $\gamma$ as $10^{-3}$ in \eq~\ref{loss}. Hyper-parameter experiments can be found in Sec.~\ref{hyper_p_exp}. 

In order to make our comparison fair, for the baseline model, its hyper-parameter settings were configured through the same method as \xxx does.

\textbf{Metrics.} In recommendation systems, items can be classified as relevant or irrelevant for a given user. We use the AUC (Area Under the Curve) score to evaluate how effectively the model can distinguish between these two classes of items. A higher AUC score indicates that the model can differentiate between relevant and non-relevant items for users in a more effective way, and have stronger capability of recommending items that users find interesting and relevant. We used CTR (Click-Through-Rate) in online experiments, and the CTR measures the ratio of the number of clicks on recommended items to the number of items displayed. A higher CTR indicates that users find the recommended items more relevant and are more likely to click on them. Furthermore, the Gini coefficient is utilized to measure the uniformity of model improvement across different scenes. A lower Gini coefficient indicates less interference by differences between scenes, better generalization performance, and better suitability for application in cold-start scenes.

\subsection{Experimental Results}

The following part mainly introduces the experimental setup and
analyzes the comparison among our \xxx model and other single-scene models (SSM) and static multi-scene models (SMSM) in the recommendation datasets of Meituan and Taobao.

\begin{table}[t]
  \caption{Comparison of AUC of each model in 10 randomly selected cold-start scenes.}
  \small
  \centering
  \begin{tabular}{c|c|c|c|c|c|c|c}
  \hline
            & \textbf{\xxx}    & MMoE    & PLE     & HMoE    & STAR    & PEPNet  & HiNet   \\ \hline\hline
  \#.1      & \textbf{0.7524}  & 0.7476  & 0.7508  & 0.7487  & 0.7432  & 0.7469  & 0.7458  \\
  \#.2      & \textbf{0.7576}  & 0.7322  & 0.7128  & 0.7306  & 0.7312  & 0.7418  & 0.7391  \\
  \#.3      & \textbf{0.8114}  & 0.7924  & 0.7753  & 0.7788  & 0.7808  & 0.7797  & 0.7815  \\
  \#.4      & \textbf{0.7644}  & 0.7508  & 0.7308  & 0.7504  & 0.7471  & 0.7493  & 0.7459  \\
  \#.5      & \textbf{0.7714}  & 0.7704  & 0.7703  & 0.7693  & 0.7626  & 0.7659  & 0.7680  \\
  \#.6      & \textbf{0.7541}  & 0.7531  & 0.7535  & 0.7512  & 0.7492  & 0.7529  & 0.7514  \\
  \#.7      & \textbf{0.7409}  & 0.7248  & 0.7142  & 0.7404  & 0.7403  & 0.7399  & 0.7388  \\
  \#.8      & \textbf{0.8178}  & 0.7824  & 0.7866  & 0.7606  & 0.7843  & 0.7802  & 0.7851  \\
  \#.9      & \textbf{0.8060}  & 0.7769  & 0.7836  & 0.7748  & 0.7755  & 0.7768  & 0.7766  \\
  \#.10     & \textbf{0.8283}  & 0.7753  & 0.7633  & 0.7736  & 0.7874  & 0.7894  & 0.7890  \\ \hline
  All       & \textbf{0.7869}  & 0.7676  & 0.7638  & 0.7587  & 0.7552  & 0.7591  & 0.7579  \\ \hline
  \end{tabular}
  \label{detail_experiment}
\end{table}

\textbf{SSM Experiments.} This experiment is first based on classical SSM, including DNN~\cite{covington2016deep}, MMoE~\cite{ma2018modeling}, and PLE~\cite{tang2020progressive}. DNN is a single-scene and single-task model. As shown in Table~\ref{experiment}, our model has significantly improved AUC compared to DNN in both datasets, especially for the recommendation effect of new scenes in Meituan's dataset.
MMoE and PLE are all single-scene and multi-task models. In this experiment, we added two objectives, click prediction and order prediction, for Dataset-1. In contrast, we conducted single-objective prediction for the Taobao dataset due to only one objective provided. In addition, to be consistent with the industrial application strategy, the SSM model uses all samples from various scenes for training and incorporates scene IDs as features. However, no multi-scene model structure optimization has been performed. The experimental results also prove that our model outperforms the baseline models in new and old scenes.

\textbf{SMSM Experiments.} To verify the effectiveness of \xxx compared to existing state-of-the-art SMSMs, we trained the HMoE~\cite{li2020improving}, STAR~\cite{sheng2021one}, PEPNet~\cite{chang2023pepnet} and HiNet~\cite{zhou2023hinet} and then conducted comparative experiments. According to the definition mentioned earlier, both of these models belong to the static multi-scene model, which is suitable for multiple fixed scenes with stable traffic, and therefore contradicts the definition of dynamic multi-scene. To solve this problem, we adopted a standard solution in industrial applications: clustering scenes based on their attributes and treating the resulting cluster of new and old scenes as a sizeable stable scene. The experimental results showed that \xxx still achieved the best performance.

\textbf{Sub-scenes Experiments.} In this context, a sub-scene refers to an individual scene within each source in Table~\ref{dataset_intro}. We randomly selected 10 sub-scenes from the test set and tested the effect comparison of different models, as shown in Table~\ref{detail_experiment}. It can be seen that \xxx achieves the highest AUC in each sub-scene.

In summary, \xxx has shown advantages in solving dynamic multi-scene problems compared to other widely-used single-scene and multi-scene models. This further proves that \xxx is theoretically practical and widely applicable.

\subsection{Hyperparameters Experiments} \label{hyper_p_exp}

To illustrate the impact of hyperparameters on the experimental results, we conducted relevant experiments based on dataset-1.

\begin{table}[t]
    \centering
    \small
    \caption{The experimental results of different $cc$ threshold.}
    \begin{tabular}{c|c|c|c}
    \hline
    $cc$ threshold           & $\pm$0.1   & $\pm$0.05        & $\pm$0.01  \\ \hline\hline
    Number of key features   & 4          & 13               & 68         \\ \hline
    AUC                      & 0.7849     & \textbf{0.7860}  & 0.7851     \\ \hline
    \end{tabular}
    \label{cc_th}
\end{table}

\textbf{Hyperparameters of SRG.}\label{hp_srg_cc} We tested the experimental results of constructing SRG by filtering features according to different correlation coefficients ($cc$) thresholds (Table.\ref{cc_th}). It can be found from the experimental results that a reasonable threshold can filter out noise features and select as many effective features as possible to improve the model performance. The empirical threshold is $0.05$, which can also be experimented with and adjusted according to specific business data.

\begin{table}[t]
  \centering
  \small
  \caption{Model under different $\alpha$ ($\beta=0.001, \gamma=0.001$).}
  \begin{tabular}{c|c|c|c}
  \hline
           & $\alpha=2$ & $\alpha=1$         & $\alpha=0.5$   \\ \hline\hline
  AUC      & 0.7859     & \textbf{0.7860}    & 0.7839         \\ \hline
  \end{tabular}
  \label{alpha_exp}
\end{table}

\begin{table}[t]
  \centering
  \small
  \caption{Model under different $\beta$ ($\alpha=1, \gamma=0.001$).}
  \begin{tabular}{c|c|c|c|c}
  \hline
              & $\beta=0.0001$   & $\beta=0.001$       & $\beta=0.01$   & $\beta=0.1$    \\ \hline\hline
  AUC         & 0.7858           & \textbf{0.7860}     & 0.7855         & 0.7850         \\ \hline
  \end{tabular}
  \label{beta_exp}
\end{table}

\begin{table}[t]
  \centering
  \small
  \caption{Model under different $\gamma$ ($\alpha=1, \beta=0.001$).}
  \begin{tabular}{c|c|c|c|c}
  \hline
          & $\gamma=0.0001$   & $\gamma=0.001$      & $\gamma=0.01$  & $\gamma=0.1$   \\ \hline\hline
  AUC     & 0.7859            & \textbf{0.7860}     & 0.7856         & 0.7852         \\ \hline
  \end{tabular}
  \label{gamma_exp}
\end{table}

\begin{table*}[ht]
  \centering
  \small
  \caption{Ablation experiments without (w/o) structures based on Meituan's Dataset. The statistical significance of \xxx's performance improvement over its alternative versions has been validated by the Friedman test.}
  \begin{tabular}{c|c|c|c|c|c|c}
  \hline
              & \xxx            & w/o SRG & w/o AEM & w/o CFR & w/o $Loss_{var}$ & w/o $Loss_{cos}$ \\ \hline\hline
  AUC of all scenes        & \textbf{0.7860} & 0.7805  & 0.7819  & 0.7840  & 0.7838           & 0.7841           \\ \hline
  AUC of cold-start scenes & \textbf{0.7799} & 0.7732  & 0.7749  & 0.7771  & 0.7767           & 0.7774           \\ \hline
  \end{tabular}
  \label{ablation_t}
\end{table*}

\begin{table}[t]
    \centering
    \small
    \caption{Model under different temperature coefficient $\tau$.}
    \begin{tabular}{c|c|c|c|c|c}
    \hline
           & $\tau=1$ & $\tau=0.1$ & $\tau=0.01$ & $\tau=0.001$ & $\tau=0.0001$ \\ \hline\hline
    AUC    & 0.7821   & 0.7838     & 0.7854      & 0.7860       & 0.7860        \\ \hline
    \end{tabular}
    \label{tau_exp}
\end{table}

\begin{table}[t]
    \centering
    \small
    \caption{Model performance under different expert number of AEG ($N_{s}=10$). The inference time refers to the time consumption of each sample, measured in milliseconds.}
    \begin{tabular}{c|c|c|c|c|c|c}
    \hline
    $N_{a}$           & 3      & 5      & 8      & 10     & 13      & 15      \\ \hline\hline
    AUC               & 0.7802 & 0.7831 & 0.7845 & 0.7860 & 0.7863  & 0.7863  \\ \hline
    Inference time    & 0.1641 & 0.1670 & 0.1710 & 0.1741 & 0.1785  & 0.1814  \\ \hline
    \end{tabular}
    \label{f_num_exp}
\end{table}

\begin{table}[t]
    \centering
    \small
    \caption{Model performance under different expert number of SEG ($N_{a}=10$). The inference time refers to the time consumption of each sample, measured in milliseconds.}
    \begin{tabular}{c|c|c|c|c|c|c}
    \hline
    $N_{s}$           & 3      & 5      & 8      & 10     & 13      & 15      \\ \hline\hline
    AUC               & 0.7815 & 0.7839 & 0.7849 & 0.7860 & 0.7862  & 0.7863  \\ \hline
    Inference time    & 0.1651 & 0.1677 & 0.1714 & 0.1741 & 0.1781  & 0.1809  \\ \hline
    \end{tabular}
    \label{s_num_exp}
\end{table}

\textbf{Hyperparameters of Loss Functions.} Regarding the hyperparameter selection of loss functions, the experimental outcomes for $\alpha$, $\beta$, and $\gamma$ are presented in Table~\ref{alpha_exp}, Table~\ref{beta_exp}, and Table~\ref{gamma_exp} (Section.\ref{sec:method:loss} for details). Given that $Loss_{ce}$ plays a pivotal role in model optimization, it is advisable to set $\alpha$ to a relatively substantial value. Conversely, as both $Loss_{cos}$ and $Loss_{var}$ serve as auxiliary components in the training process, it is recommended to keep $\beta$ and $\gamma$ small-valued.

\textbf{Hyperparameters of $Dics$.} Based on the experimental findings for temperature coefficients $\tau$ of $Dics$ as shown in Table~\ref{tau_exp}, it is evident that excessively high temperature coefficients result in a decline in performance, leading to reduced diversity among various scenes in AEG. Conversely, referencing Eq.~\ref{dics}, overly small temperature coefficients introduce discontinuities in the curve of the $Dics$ function, thereby destabilizing the training process. Hence, a temperature coefficient value of around 0.001 can be selected and appropriately increase the value of $Loss_{var}$.

\textbf{Hyperparameters of AEM.} The experiments regarding the number of experts in AEG ($N_{a}$) and SEG ($N_{s}$) can be found in Table~\ref{f_num_exp} and Table~\ref{s_num_exp}. It's worth noting that the impact of adding experts is most pronounced when $N_{a}$ and $N_{s}$ are relatively small. However, as these values grow larger, the model's computational efficiency decreases, and the gains in performance become less significant. Therefore, it is advisable to strike a reasonable balance between effectiveness and efficiency.

\begin{table*}[h]
  \caption{Online Experimental Results.}
  \small
  \centering
  \begin{tabular}{c|c|c|c|c|c|c|c}
  \hline
                                       & Sub Scene \#1 & Sub Scene \#2 & Sub Scene \#3 & Sub Scene \#4 & Sub Scene \#5 & Sub Scene \#6 & Gini   \\ \hline\hline
  \#Exposure (1 day)                     & $16K$         & $33K$         & $20K$         & $65K$         & $7K$          & $48K$         & -      \\
  \#CTR                                & 21.97\%       & 9.46\%        & 0.36\%        & 1.39\%        & 27.18\%       & 0.72\%        & -      \\ \hline
  Baseline / Default                   & +50.60\%      & +47.20\%      & +38.19\%      & +53.51\%      & +33.35\%      & +46.53\%      & 0.0858 \\
  Ours / Default                       & +53.91\%      & +49.69\%      & +41.56\%      & +55.07\%      & +37.70\%      & +48.57\%      & 0.0727 \\
  Ours / Baseline                      & +6.55\%       & +5.29\%       & +8.83\%       & +2.92\%       & +13.07\%      & +4.40\%       & 0.2651 \\ \hline
  \end{tabular}
  \label{online_experiment}
\end{table*}

\subsection{Ablation Study and Analysis}

The results of the ablation experiments are shown in Table~\ref{ablation_t}. Firstly, we tested the impact of SRG on the model performance. SRG is mainly responsible for introducing prior knowledge of similar scenes to the model, which is the theoretical basis of \xxx. After removal, other affected structures must be randomly initialized and uniformly distributed. This structure significantly impacts the recommendation performance (the AUC decreased from 0.7860 to 0.7805 after removal). Secondly, we conducted ablation experiments on the AEM structure. This component is responsible for learning how to design the model structure and extracting exclusive and shared information for each scene. The control group for this experiment used uniformly distributed experts instead of the original dynamic allocation in AEG. The results showed a certain degree of degradation in AUC (from 0.7860 to 0.7819). Thirdly, the CFR was masked, and the AUC dropped to 0.7840. Finally, we performed ablation experiments on $Loss_{var}$ and $Loss_{cos}$. After adding $Dics(\cdot)$, AEG can dynamically allocate experts, and $Loss_{var}$ further enhances the discrimination of the allocation. The experimental results showed that removing the two types of losses decreased 22 BP (Basic Point) and 19 BP in the model AUC, respectively.

In the research field of recommendation, improving a model to achieve higher AUC than state-of-the-art recommendation approaches is generally recognized to be highly challenging in practice~\cite{sheng2021one,zhou2018deep,zhou2019deep}. The comparative experiments in this paper were repeated 10 independent times for validation, and the results were statistically significant at the $0.05$ level (Friedman test), indicating confidence in the effectiveness of \xxx.

\subsection{Application in Practice} \label{app_in_prac}

SwAN has been deployed in the online recommendation system to validate its practical effectiveness and component efficacy.

\textbf{Online Application Performance.} Besides the importance in theory, recommendation plays a pivotal role in commercial situations. To further demonstrate the superior performance of \xxx in practical applications, we deployed it in the online recommendation system of Meituan's catering business with an average daily user level of millions, which has typical dynamic multi-scene characteristics, and conducted an A/B test with 20\% of the traffic over a period of two months. More than 200 new online scenes were added during the experiment, and the total number exceeded 400. The experimental results showed that \xxx achieved a 5.64\% increase in the CTR index compared to the best baseline model (PLE) and a 5.19\% increase in daily order volume proportion after full traffic promotion.

In addition, we randomly selected 6 new scenes online and calculated the CTR improvements of both \xxx and baseline models relative to the default ranking method (Table~\ref{online_experiment}):

First, the baseline models and \xxx have significant CTR enhancements relative to the default ranking. However, \xxx has a smaller Gini coefficient for the improvement ratio between scenes, demonstrating superior stability. We also calculated the Gini coefficient of the improvement ratio of \xxx relative to the baseline model in each scene, which is 0.2651 (0.2351 in all scenes). This value proves that the improvement of \xxx relative to the baseline model is evenly distributed among different scenes instead of only focusing on large scenes and ignoring small ones.

Second, combining each scene's daily exposure samples, we found that \xxx has a more significant improvement ratio than the baseline model in small scenes, demonstrating its better generalization ability and competence to optimize cold-start problems for scenes with sparse user behaviors.

\begin{figure}[t]
  \centering
  \includegraphics[scale=0.6]{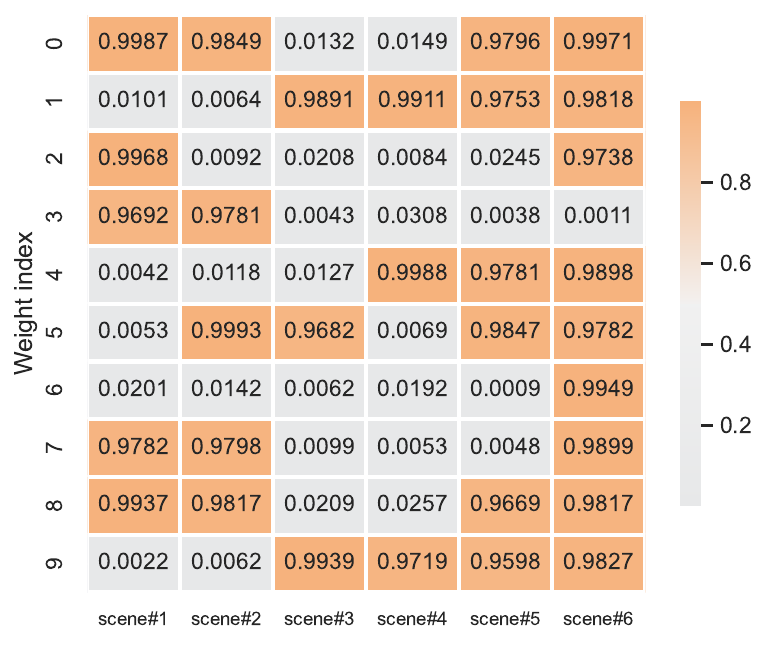}
  \caption{$W_{k}$ (10 selected) in the AEG calculated from samples of 6 randomly chosen scenes. Each column represents 10 $W_{k}$ calculated from a sample.}
  \label{weights_show}
\end{figure}

\textbf{Visualization of the Expert Selector.} To demonstrate the effectiveness of information transfer in Expert Selector, we randomly selected 6 scenes and one sample from each scene. The $W_{k}$ values in AEG calculated from these samples are shown in Fig.~\ref{weights_show}.

Firstly, it can be seen that the expert selection in different scenes is significantly different, indicating that Expert Selector can distinguish between different scenes. Secondly, from the figure, it can be observed that scene\#1 and scene\#2 exhibit a high degree of similarity in the selection of experts. Upon examining the actual scene data, we find that both scenes primarily focus on selling afternoon tea. This further confirms that the selection of experts is related to the similarity of the actual scenes. Finally, it can be seen that the model increases the number of selected experts in scene\#6 due to the diverse item types, which strengthens its generalization performance.

\subsection{Model Complexity} \label{model_coplexity}

\begin{table}[ht]
  \caption{Time and space complexity of models on the Meituan dataset. The inference time refers to the time consumption of each sample, measured in milliseconds.}
  \small
  \centering
  \begin{tabular}{c|c|c|c}
  \hline
  Model         & Params ($\times 10^{7}$) & Training time  & Inference time  \\ \hline\hline
  DNN           & 1.90                     & 142 mins       & 0.1389          \\
  MMoE          & 1.99                     & 164 mins       & 0.1442          \\
  PLE           & 2.06                     & 176 mins       & 0.1675          \\
  HMoE          & 2.06                     & 168 mins       & 0.1739          \\
  STAR          & 2.18                     & 178 mins       & 0.1808          \\
  PEPNet        & 2.39                     & 188 mins       & 0.1889          \\
  HiNet         & 2.27                     & 182 mins       & 0.1865          \\
  \xxx          & 2.21                     & 173 mins       & 0.1741          \\ \hline
  \end{tabular}
  \label{model_complexity}
\end{table}

All experiments were conducted on NVIDIA Tesla A100 GPU, 80G RAM, and Intel(R) Xeon(R) Gold 5218 CPU servers. Table~\ref{model_complexity} shows that \xxx maintains a reasonable number of parameters and prediction time. Generally, we retrieve online data in real-time for training and update the model approximately every half hour.

%% file: sections/conclusion.tex
In this paper, we propose the \xxx model, a novel approach to addressing the cold-start problem in Multi-scene Recommendation (MSR) systems. The proposed model overcomes the limitations of traditional MSR approaches by directly and significantly enhancing the performance of newly-arrived scenes through online prediction. The unique architecture of the \xxx model, which combines the Scene Relation Graph (SRG), Similarity Attention Network (SAN), and Adaptive Ensemble-experts Module (AEM), enables it to capture graph-structured similarities between scenes, understand user behavior transitions, and identify shared logic among different scenes. Our extensive evaluation of \xxx on both public and Meituan industrial datasets demonstrate the superiority of the \xxx model over existing MSR approaches, providing more accurate and high-quality recommendations in a dynamic and adaptable manner. Furthermore, \xxx has been deployed in the online catering recommendation service of Meituan, which serves millions of daily customers, and has achieved a significant improvement in CTR (Click-Through Rate) index. This work represents a significant step forward in developing efficient and adaptable recommendation systems, particularly in the context of a rapidly evolving E-commerce landscape.